# Complexity Reduction of Learned In-Loop Filtering in Video Coding

Woody Bayliss
Queen Mary University
of London *MMV*
London, United Kingdom
W.Bayliss@qmul.ac.uk

Luka Murn
British Broadcasting
Corporation *R&D*
London, United Kingdom
Luka.Murn@bbc.co.uk

Ebroul Izquierdo
Queen Mary University
of London *MMV*
London, United Kingdom
Ebroul.Izquierdo@qmul.ac.uk

Qianni Zhang
Queen Mary University
of London *MMV*
London, United Kingdom
Qianni.Zhang@qmul.ac.uk

Marta Mrak
British Broadcasting
Corporation *R&D*
London, United Kingdom
Marta.Mrak@bbc.co.uk

*Abstract*— In video coding, in-loop filters are applied on reconstructed video frames to enhance their perceptual quality, before storing the frames for output. Conventional in-loop filters are obtained by hand-crafted methods. Recently, learned filters based on convolutional neural networks that utilize attention mechanisms have been shown to improve upon traditional techniques. However, these solutions are typically significantly more computationally expensive, limiting their potential for practical applications. The proposed method uses a novel combination of sparsity and structured pruning for complexity reduction of learned in-loop filters. This is done through a three-step training process of magnitude-guided weight pruning, insignificant neuron identification and removal, and fine-tuning. Through initial tests we find that network parameters can be significantly reduced with a minimal impact on network performance.

*Keywords — neural networks, video coding, in-loop filtering, quality enhancement, pruning, sparsity*

## I. INTRODUCTION

Recent research efforts show that learned tools, such as deep Neural Networks (NNs), can be successfully applied to improve the performance of video compression algorithms. For example, the upcoming Versatile Video Coding (VVC) standard already includes a tool derived from learning-based techniques [1]. Furthermore, methods based on deep learning have been used as intra/inter-prediction [2-5], quantisation, entropy coding [6] and loop filtering tools [7].

When compressing a video at higher quantization rates, artifacts can occur due to poor frame reconstruction. In-loop filtering is a well-known video compression method applied at the end of the coding process that aims to reduce or completely remove artifacts. VVC implements three filters, namely the Deblocking Filter (DBLF), the Sample Adaptive Offset filter (SAO), and the Adaptive Loop Filter (ALF).

The loop filtering process can be expanded with new filters that apply deep learning techniques, as demonstrated with Attention-Based Dual-Scale CNN (ADCNN) [7]. ADCNN utilises Coding Unit (CU) maps as an attention mechanism that guides the neural network, to remove artifacts such as blocking.

ADCNN has exhibited improved compression performance over the baseline VVC. However, the encoder processing time is increased by 66% on a Central Processing Unit (CPU). The decoder processing time is increased by more than 130 times when processed by a Graphics Processing Unit (GPU), and by more than 450 times with a CPU. The substantial increase in coding running time makes this method unfeasible for practical solutions.

This paper proposes to reduce the coding complexity of learning-based methods within video coding applications by developing an automated NN pruning methodology. Pruning identifies redundant parameters from trained NNs and removes them to reduce the NN size with minimal impact on network accuracy. The presented approach inspired by ADCNN, is an initial attempt to test the feasibility of pruning NN in-loop filters. This work focuses on pruning three single-branch networks at the decoder stage, each processing separate Y, U, V components.

The rest of the paper is organised as follows. Section II introduces related work in the areas of deep learning, in-loop filtering and neural network pruning. Section III describes the proposed pruning methodology, while Section IV analyses the experimental results of the trained network applied within VVC. Section V presents conclusions.

## II. RELATED WORK

### A. In-loop Filtering

In-loop filtering is an essential operation for lossy video coding standards, as compression removes information from raw video data to reduce its bitrate. Modern video coding standards such as High-Efficiency Video Coding (HEVC) [8] and VVC [1] divide regions of an input frame into Coding Tree Units (CTUs), which further separate regions into rectangular blocks or CUs. During the encoding step, each CU is predicted with different methods, such as inter/intra-prediction, merge mode or skip mode. The final choice is selected according to the approach that minimizes the Rate-Distortion (RD) cost in the presence of quantisation. The level of quantisation is defined by a Quantisation Parameter (QP), high QP values can cause artifacts within the reconstructed frame.

Video coding artifacts are present in many forms. For example, blockiness can occur within frames on the borders of adjacent CUs. Additionally, ringing artifacts can appear in large CUs. VVC applies several filters to reduce artifacts. DBLF is used to apply deblocking and softens the borders between CUs, while SAO reduces ringing distortions [9]. A third filter, ALF/ALF-CC, is used to restore the objective quality, expressed in PSNR, lost by the application of the previous filters [9].

### B. Deep Learning Approaches for In-loop Filtering

When applied to video coding, deep learning methods for in-loop filtering often utilise residual models [7,10,11]. Residual models learn the difference, or the residual, between the output and the input of the network [12]. Learning the residual means that the network weights are sparser than if required to learn the entire frame reconstruction, and it is therefore an easier task to approximate. Additionally, by stacking multiple repeated residual layers, weights in each subsequent layer become more specialised [13], improving model performance.

Challenging deep learning tasks such as image restoration, denoising and super-resolution share similar properties with loop filtering problems, as they also aim to restore and enhance frame quality. Therefore, Convolutional Neural Networks (CNNs) used for these tasks have also been successfully applied to in-loop filtering. The seminal Super-Resolution Convolutional Neural Network (SRCNN) [14] extracts features from the input and maps them to high resolution patch representations to improve the resolution of JPEG encoded images. SRCNN was extended in size [15] and applied to picture compression artifacts, spatial temporal information further improved model performance in [2].

Inspired by the SRCNN structure, an In-loop Filtering CNN (IFCNN) was proposed in [16] and used as a replacement for the DBLF and SAO filters in HEVC, achieving -4.8 % Bjøntegaard-Delta rate (BD-rate) [17] savings under the All-Intra (AI) configuration. A residual network called Enhanced Deep Convolutional Neural Network (EDCNN) achieved a -6.45 % BD-rate reduction on average for all HEVC configurations [10]. A CNN architecture for post-processing and in-loop filtering called MFRNet was introduced in [11] and achieved up to -5.1 % coding gains when integrated within VVC. However, it increased the decoder running time more than 80 times. Combining wide activated models [18] and squeeze-excitation models [19] resulted in an Attention-Based Dual-Scale CNN (ADCNN) [7] which, when implemented in VVC, achieved -6.54%, -13.27% and -15.72% BD-rate savings over traditional filters for the Y, U and V channels, respectively. This model is attention based, with the attention input being the CU map extracted from the decoder, selected filtered blocks are signalled with flags in the bit stream.

*C. Neural network pruning*

Neural network pruning identifies redundant components within a trained neural network. Learned parameters are zeroed out [20] or removed from the model entirely [21].

Sparsity pruning is applied on pre-trained networks and aims to reduce the number of active neuron connections. The network is sparsified by setting weight values in network layers to zero. When applied to large-scale deep NNs, sparsity pruning can remove up to 80 % of learned network connections with a minimal impact on network performance [20]. While sparsity pruning does not reduce model parameters, structured pruning physically removes neurons from neural networks [21]. Structured pruning is highly dependent on its application and the type of NN used.

For visual data applications that utilise CNNs, structured pruning can remove entire trained convolutional channels from layers of the network [22,23], significantly reducing the trained network size. The approach in [22] structurally prunes image recognition networks ConvNet, AlexNet, VGG-16 and ResNet-50 by removing channels, achieving speed ups in computation up to 4 times.

Several algorithms for identifying redundant neurons in structured NN pruning have been proposed. A data-driven approach in [24] uses validation data to calculate the Average Percentage of Zeros (APoZ) for a given layer. A high APoZ indicates many redundant neurons. Filter clustering identifies similar filter groups and retains one filter per group [25]. Filters can also be removed based on the L1 norm of their weights [26] or their overall contribution to the output feature maps [27].

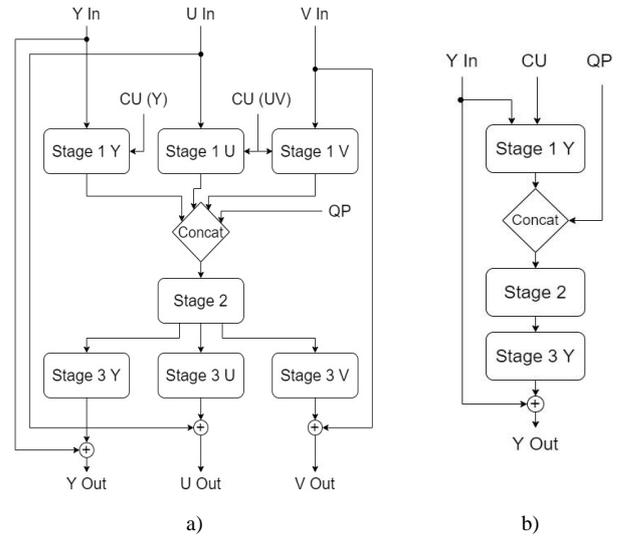

Fig. 1. a) General structure of the ADCNN model. b) Network structure for a channel after separating the Y, U, V network into three UCLF networks.

III. METHODOLOGY

The ADCNN architecture, illustrated in Fig. 1a. provides significant improvements when compared to traditional filters. The basic structural blocks used in ADCNN are used for constructing the separate models of our approach. Spatial attention was removed from the blocks of ADCNN, as the proposed pruning method is only applicable on single-branch networks.

The resulting model was named Uni-Component Loop Filter (UCLF), shown in Fig. 1b. for the Y channel. Two models for the U and V channels also have the same architecture as the one for Y. The architectures are separated into three stages, each stage consists of generalised residual and non-residual blocks. The residual block structure is displayed in Fig. 2a. with two convolutional layers (2DConv) of 3-by-3 kernel sizes followed by two dense layers. Non-residual blocks follow the same structure as residual ones, without the input being added to the output. Each model was individually pruned using a novel combination of sparsity and structured pruning.

The NN-based filter in ADCNN is integrated within VVC as a switchable filter, meaning that the encoder can select either conventional filters (DBLF, SAO, ALF) or the learned filter to obtain a reconstructed rectangular block within the currently coded frame. The filter choice is decided according to the RD cost calculation. However, our approach applies learned in-loop filters directly on each blocky video frame, rather than through a switchable implementation.

*A. Pruning Algorithm*

The proposed pruning method is detailed in Algorithm 1 (Subsection III-A). For a pretrained neural network **T** sparsity pruning (Subsection III-B) is applied to each prunable layer. The validation set is passed through the intermediary model, and its activation maps are then used to identify redundant channels (Subsection III-C). A pruned network **P** is obtained from **T** by iteratively analysing the activation maps and removing channels from each layer that have values lower than a set pruning threshold (Subsection III-D). **P** is then re-trained for a defined number of optimization epochs to retain performance.

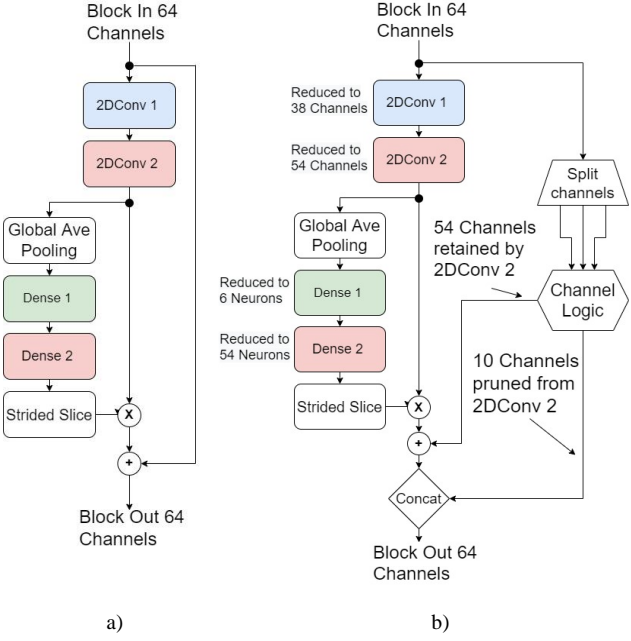

Fig. 2. a) General structure of the residual blocks in each stage. b) An example of the proposed structured pruning method applied to a residual block.

Following the re-training step, the accuracy, and the number of parameters of **P** are recorded and compared to pre-set thresholds. The pruning algorithm stops if any of the constraints are satisfied, once a constraint is satisfied **P** from the previous iteration is returned as the final, pruned model. Otherwise, **T** = **P** and the process repeats.

**Algorithm 1.**

PROPOSED PRUNING ALGORITHM

**Input**: Pretrained neural network **T**, number of parameters *num_par*, list of prunable layers *pl*, training samples *x*, validation samples *v*, sparsity_threshold *st* (Sec. III.B), channel_threshold *ct* (Sec. III.C), number of optimization epochs *train_epochs* (Sec. III.A), accuracy_threshold *at* and pruning_threshold *pt* (Sec. III.D)

**Output**: Pruned neural network **P**

while *True:*

   for *layer* in **T**:

      if *layer* in *pl*:

         *model* = apply_sparsity_pruning(*layer*, **T**, *st*)

   *chan_to_remove* = identify_redundant_channels(*v*, *model*, *ct*)

   **P** = apply_structured_pruning(*model, chan_to_remove*)

   **P**.train(*x, train_epochs*)

   if **P**.accuracy < *at* or **P**.*num_par* / **T**.*num_par* < 1 - *pt*:

      **P** = **T**

      **break**

   else

      **T** = **P**

TABLE I   NUMBER OF PRUNABLE CHANNELS AT EACH NETWORK STAGE

| Stage | Block type | No. of blocks | Prunable channels per block | | | |
|---|---|---|---|---|---|---|
| | | | 2DConv 1 | 2DConv 2 | Dense 1 | Dense 2 |
| 1 | Non-res. | 2 | 48 | -- | 8 | -- |
| | Residual | 1 | 48 | 32 | 8 | 32 |
| 2 | Residual | 5 | 96 | 64 | 16 | 64 |
| 3 | Non-res. | 2 | 48 | -- | 8 | -- |
| | Residual | 1 | 96 | 64 | 16 | 64 |

*B. Sparsity Pruning*

Sparsity pruning sets weight values within a layer to zero according to a specified sparsity threshold. For the pretrained UCLF network from Fig. 1b, it is applied on both residual and non-residual blocks in all three stages. To retain network performance sparsity was only applied to specific layers within the model. Additionally, the first and last layer of the network are not pruned. The total number of prunable convolutional channels and dense units is presented in Table I. The values listed in Table I indicate the initial numbers of filters and dense units in a UCLF network, these numbers will reduce as pruning progresses.

*C. Insignificant Channel Identification*

Sparsity pruning does not guarantee that all weights associated with a channel will be zero. When applied to a certain layer, magnitude-based weight pruning sparsifies the entire layer across channels, rather than within each channel individually. In convolutional layers, filters are considered as channels. In dense layers, neurons are also considered as channels.

To identify which channels in a layer can be removed from the UCLF network, a data-driven approach is adopted. The validation set is used as an input to the network. Neuron activation and filter activation maps are stored at each prunable layer. The stored values are then averaged over the entire validation set. If the average channel value is below the channel threshold, then that channel is marked for removal.

*D. Structured Pruning*

Structured pruning removes channels that were identified as insignificant for a specific UCLF network. For channels within convolutional and dense layers, this means removing all associated weight and bias information.

An example of a pruned residual block from a UCLF network is presented in Fig. 2b. As each stage of the network consists of stacked generalised blocks from Fig. 2a. the dimensionality at the input and the output must be retained. In this example, the input is set to 64 channels. The first convolutional layer, marked in blue, and the first dense layer, marked in green, can be pruned without restrictions. The second convolutional layer and the second dense layer need to have an equal number of channels, as they are multiplied together. Therefore, when a channel is removed from the second convolution it must also be removed from the second dense layer. The pruned channels are then added to the corresponding channels from the input to the block. Finally, these are concatenated to the rest of the input channels and constitute the output of the residual block.

TABLE II Coding performance of proposed approach in VTM 7.0 for the AI configuration, tested on CTC sequences

| Class | UCLF before pruning | | | | | | UCLF after pruning | | | | | | Time Reduction (50 frames) |
|---|---|---|---|---|---|---|---|---|---|---|---|---|---|
| | BD-rate [%] | | | BD-PSNR [dB] | | | BD-rate [%] | | | BD-PSNR [dB] | | | |
| | Y | U | V | Y | U | V | Y | U | V | Y | U | V | |
| B | -3.53 | -3.82 | -2.58 | 0.13 | 0.07 | 0.06 | -3.07 | -4.75 | -3.90 | 0.12 | 0.09 | 0.09 | 31% |
| C | -4.68 | -5.16 | -2.37 | 0.29 | 0.19 | 0.21 | -4.52 | -5.91 | -5.30 | 0.28 | 0.22 | 0.21 | 36% |
| D | -6.57 | -7.58 | -8.68 | 0.47 | 0.30 | 0.35 | -6.43 | -7.93 | -9.03 | 0.46 | 0.32 | 0.37 | 44% |
| E | -5.13 | -2.35 | -2.12 | 0.25 | 0.08 | 0.06 | -5.09 | -3.89 | -3.49 | 0.25 | 0.14 | 0.11 | 59% |
| Average | -4.98 | -4.73 | -3.94 | 0.29 | 0.16 | 0.17 | -4.78 | -5.62 | -5.43 | 0.28 | 0.19 | 0.20 | 42% |
| #Par | Y: 879,681; U: 879,681; V: 879,681 | | | | | | Y: 667,265; U: 293,811; V: 116,972 | | | | | | |
| Time [s] | Y: 285; U: 120; V: 120 | | | | | | Y: 243; U: 84; V: 76 | | | | | | |

## IV. RESULTS

In this section, a description of the dataset, training and testing configuration for UCLF networks is presented, followed by experiments that validate the proposed model pruning approach.

### A. Dataset, model training and testing configuration

UCLF networks for Y, U and V channels are trained on the DIV2K dataset [28], which contains 800 high definition high resolution images for training and 100 images for validation purposes. Images are separately encoded with VVC Test Model (VTM) 7.0 [29] under the All-Intra (AI) configuration at 4 QPs levels, 22, 27, 32, 37. Common Test Conditions (CTC) as defined by JVET [30] are modified to disable the three in-loop filters within VVC, to obtain blocky images as network inputs. Additionally, CU map information was extracted for each encoded image. Image patches of size $48 \times 48$ are used for training the Y network, while $24 \times 24$ patches are used for the U and V networks.

The individual Y, U and V models are trained with the Mean Absolute Error (MAE) loss function, Adam optimizer and a learning rate set to 0.001. The trained, individual models are then pruned according to Algorithm 1. Sparsity pruning is configured to achieve 80 % sparsity for the intermediary model. A channel value threshold of 0.001 is used. The models are trained and tested on an NVIDIA Quadro RTX 5000 GPU.

After the pruning process is finished, the learned in-loop filters are tested on CTC video sequences, where each class represents a set of sequences of same spatial resolution. All sequences were processed in the same manner as the training dataset. The obtained blocky videos are filtered by the pruned networks and then compared to videos encoded by the baseline VTM 7.0 anchor.

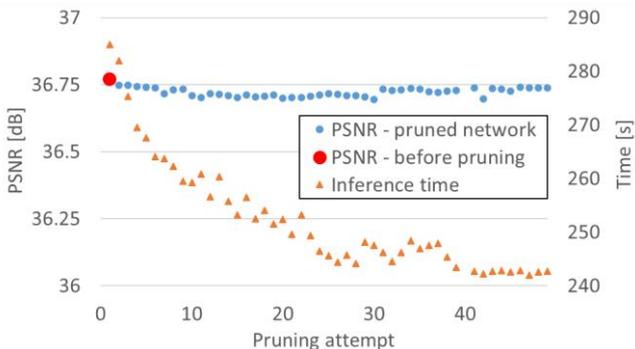

Fig. 3. Pruning of UCLF Y component network. Each pruning attempt represents one pruning loop of Algorithm 1. Results are displayed as average PSNR and total inference time for the validation dataset.

### B. Network performance during pruning

The performance of a pre-pruned UCLF network during the pruning process is reported in Fig. 3. Two metrics, PSNR and inference time, are measured during each pruning loop. The pre-trained network for the Y channel reports an average PSNR of 36.77 dB on the validation dataset and requires 285 seconds to process the dataset on a GPU.

With each iteration, inference time decreases, while the PSNR remains stable. Compared to the original pre-trained network, the final pruned network produces a PSNR value lower by 0.04 dB and a reduction of 15 % in processing time.

The coding performance of unpruned and pruned UCLF networks for in-loop filtering is compared in Table II. A significant reduction in the number of parameters for each network is observed, with a slight decrease in BD-rate for the Y channel. However, the pruned networks for the U and V channels exhibit higher BD-rates and increase the overall BD-PSNR more than the unpruned baselines.

It must be noted that the reduction in parameters only provides a general measure of network complexity, whilst the inference time shows the real benefits of pruning a network with the proposed approach. The pruned network for the V channel has nearly 87% less parameters than its unpruned counterpart but displays 37% lower inference time.

On average, the pruned networks increase the BD-PSNR by 0.05 dB more than unpruned ones, while processing the video sequences 23 % faster. The results suggest that the original models may be over-parameterized and contain redundant information that can be removed through pruning.

## V. CONCLUSIONS

An initial approach for reducing complexity of learned in-loop filters has been presented. The approach combines sparsity pruning and structured pruning to remove redundant parts of a neural network without heavily impacting its performance.

Experimental results show that this method can reduce the number of parameters of in-loop filtering networks by as much as 87 % and improve inference time by up to 59 %. Our results show this method has minimal impact on PSNR, and, in some cases, PSNR performance can improve.

The presented method has the potential to reduce the size of neural networks used in video coding, making them applicable for practical applications. Future work will focus on improvements to redundant neuron identification, pruning of multi-branch networks, and application of the method to other neural networks used as video compression tools. These improvements will allow for direct a comparison with other similar methods.

ACKNOWLEDGEMENTS

This work was supported by the UK Engineering and Physical Sciences Research Council (EPSRC) grant 2246465 for Queen Mary University of London, Multimedia and Vision Group (MMV), and the British Broadcasting Corporation (BBC).